\documentclass[pre]{revtex4}

\usepackage{graphics}
\usepackage{graphicx}

\begin{document}

\vsize 24cm

\rightline{\small{{\it Modelling Critical \& Catastrophic Phenomena in 
Geoscience: A Statistical Physics Approach}}}

\rightline{\small {Eds. P. Bhattacharyya \&
B. K. Chakrabarti, Springer, Heidelberg (2005), pp. 3-26}}

\title{Statistical Physics of Fracture and Earthquake}

\author{Bikas K. Chakrabarti}

\affiliation{Theoretical Condensed Matter Physics Division and \\Centre for 
Applied Mathematics and Computational Science,\\ Saha Institute of 
Nuclear Physics,\\ 1/AF Bidhannagar, Kolkata 700064, India.}

\email{bikask.chakrabarti@saha.ac.in}

\begin{abstract}
This chapter introduces the fracture nucleation process, their
(extreme) statistics in disordered solids, in fiber bundle models, and in the
two fractal overlap models of earthquake.
\end{abstract}

\maketitle

\section{Introduction}

\subsection{Models of Fracture in Disordered Solids and 
Statistics}

If one applies tensile stress on a solid, the solid elongates and gets strained.
The stress ($\sigma$) - strain ($\epsilon$) relation is linear for small 
stresses (Hooke's law) after which nonlinearity appears, in most cases. 
Finally at a critical stress $\sigma_f$, depending on the material, 
amount of disorder and the specimen
size etc., the solid breaks into pieces; fracture occurs. In the case of brittle
solids, the fracture occurs immediately after the Hookean linear region, and
consequently the linear elastic theory can be applied to study the essentially
nonlinear and irreversible static fracture properties of brittle solids
\cite{bkc-Lawn:1993}. 
With extreme perturbation, therefore, the mechanical or electrical properties
 of solids tend to get destabilised and failure or breakdown occurs. In
fact, these instabilities in the solids often nucleate around disorder, which
then plays a major role in the breakdown properties of the solids. The
growth of these nucleating centres, in turn, depends on various statistical 
properties of the disorder, namely the scaling properties of percolating 
structures, its fractal dimensions, etc. These statistical properties of
disorder induce some scaling behaviour for the breakdown of the disordered 
solids \cite{bkc-Chakrabarti:1997,bkc-Sahimi:2003}.

Obviously with more and more random voids, the linear response of
e.g. the modulus of elasticity $Y$ (say, the Young's modulus) of the solid
decreases. So also does the breaking strength of the material: the 
fracture
strength $\sigma_f$ of the specimen. For studying most of these mechanical
(elastic) breakdown problems of randomly disordered solids,
one can take the lattice model of disordered solids. In these
lattice models, a fraction $p$ of the bonds (or sites) are intact, with the rest
$(1 - p)$ being randomly broken or cut. Fluctuations in the
random distribution give rise to random clusters of springs inside the
bulk, for which the statistics is well developed \cite{bkc-Stauffer:1992}, 
and one can investigate the effect of the voids or impurity clusters on the 
ultimate strength of the (percolating) elastic network of the bulk solid 
\cite{bkc-Chakrabarti:1997,bkc-Sahimi:2003}. 
One can also consider and compare the results for failure 
strength of the solids with random bond strength distribution,
as for example, in random fiber bundle models (see e.g.
\cite{bkc-Pierce:1926,bkc-Daniels:1945,bkc-Pradhan:2003}).

As is well known, the initial variations (decreases) of the linear responses
like the elastic constant $Y$ so the breakdown strengths $\sigma_f$ is
analytic with the impurity (dilution) concentration. Near the 
percolation threshold \cite{bkc-Stauffer:1992} $p_c$, up to (and at) which the solid network 
is marginally connected through the nearest neighbour occupied bonds or 
sites and below which the macroscopic connection
ceases, the variations in these quantities with $p$ are expected to become
singular; the leading singularities being expressed by the respective critical
exponents. The exponents  for the modulus of elasticity $Y \sim \Delta p^{T_e}$
$\Delta p = (p - p_c)/p_c$ are well-known and depend essentially on the 
dimension $d$ of the system (see e.g., \cite{bkc-Stauffer:1992}). 
One kind of investigation searches for the corresponding
singularities for the essentially nonlinear and irreversible properties of
such mechanical breakdown strengths for $p$ near the percolation 
threshold $p_c$: to find the exponent $T_f$ for the average fracture stress

$\sigma_f \sim \Delta p^{T_f}$ for $p$ near $p_c$.
Very often, one maps \cite{bkc-Griffith:1920} the problem of breakdown to
the corresponding linear problem (assuming brittleness up to the breaking 
point) and then derives \cite{bkc-Chakrabarti:1997} the scaling relations giving the
breakdown exponents $T_f$ in terms of the linear response exponent 
($T_e$) and other lattice statistical exponents (see next section).

Unlike that for the 'classical' linear responses of such solids, the 
extreme nature of the breakdown statistics, nucleating from the weakest point of
the sample, gives rise to a non-self-averaging property. We will discuss
(in the next section)
these distribution functions $F(\sigma)$, giving the cumulative
probability of failure of a disordered sample of linear size $L$. We show that
the generic form of the function $F(\sigma)$ can be either the 
Weibull [2] form
\begin{equation}
\label{bkc-weibull}
F(\sigma) \sim 1-\exp \left [ -L^d \left ( \frac{\sigma}{\Lambda(p)}\right )^{\frac{1}{\phi}}\right ]
\end{equation}
or the Gumbel [2] form 
\begin{equation}
\label{bkc-gumbell}
F(\sigma) \sim 1-\exp \left [ -L^d \exp \left ( \frac{\Lambda(p)}{\sigma}\right )^{\frac{1}{\phi}}\right ]
\end{equation}
where $\Lambda(p)$ is determined by the linear response 
like the elasticity of the disordered solid by some other lattice
statistical quantity etc. and $\phi$ is an exponent discussed in the 
next section.

In another kind of model for disordered systems,
a loaded bundle of fibers represents the various aspects of fracture
process through its self-organised dynamics.
The fiber bundle model
study was initiated by Pierce \cite{bkc-Pierce:1926} in the context of testing
the strength of cotton yarns. Since then, this model has been studied
from various points of view. Fiber bundles are of two classes with
respect to the time dependence of fiber strength: The `static' bundles
contain fibers whose strengths are independent of time, whereas the
`dynamic' bundles are assumed to have time dependent elements to capture
the creep rupture and fatigue behaviors. For simplicity, we will discuss
here the `static' fiber bundle models only. According to the load sharing
rule, fiber bundles are being classified into two groups: Equal load-sharing
(ELS) bundles or democratic bundles and local load-sharing (LLS) bundles.
In democratic or ELS bundles, intact fibers bear the applied load equally
and in local load-sharing bundles the terminal load of the failed
fiber is given equally to all the intact neighbors. The classic work
of Daniels \cite{bkc-Daniels:1945} on the strength of the static fiber bundles
under equal load sharing (ELS) assumption initiated the probabilistic
analysis of the model (see e.g., \cite{bkc-Pradhan:2003}). 
The distribution of burst avalanches
during fracture process is a marked feature of the fracture dynamics 
and can be observed in ultrasonic emissions during the fracture process.
It helps characterizing different physical systems along with the
possibility to predict the large avalanches. From a nontrivial probabilistic
analysis, one gets \cite{bkc-Hemmer:1992} power law distribution
of avalanches for static ELS bundles, whereas the power law exponent
observed numerically for static LLS bundles differs significantly.
This observation induces the possibility of presenting loaded fiber
bundles as earthquake models (see Sec. 3). 
The phase transition \cite{bkc-Pradhan:2003}
and dynamic critical behavior of the fracture process in such bundles
has been established through recursive formulation 
\cite{bkc-Pradhan:2003,bkc-Pradhan:2001} of the failure dynamics. 
The exact solutions \cite{bkc-Pradhan:2001}
of the recursion relations suggest universal values of the exponents
involved. Attempt has also been made \cite{bkc-Hidalgo:2001} to study
the ELS ans LLS bundles from a single framework introducing a `range
of interaction' parameter which determines the load transfer rule.

\subsection{Earthquake Models and Statistics}
The earth's solid outer crust, of about 20 kilometers in 
average thickness, 
rests on the tectonic shells. Due to the high temperature-pressure
phase changes and the consequent powerful convective flow in
the earth's mantle, at several hundreds of kilometers of depth,
the tectonic shell, divided into a small number (about
ten) of  mobile plates, has relative velocities of the order of a 
few centemeters per year \cite{bkc-Gutenberg:1954,bkc-Kostrov:1990}. 
Over several tens of years, 
enormous elastic strains develop sometimes on the earth's
crust when sticking (due to the solid-solid friction) to the
moving tectonic plate. 
When slips occur between the crust and
the tectonic plate, these stored elastic energies are 
released in 'bursts', causing the damages during the earthquakes.
Because of the uniform motion of the tectonic plates, the 
elastic strain energy stored in a portion of the crust (block),
moving with the plate relative to a 'stationary' neighbouring portion
of the crust,  can vary only due to the random strength of
the solid-solid friction between the crust and the plate. The slip
occurs when the accumulated stress exceeds the frictional force.

As in fracture (in fiber bundle model in particular), the observed 
distribution of the elastic energy release in various earthquakes 
seems to follow a power law. The number of earthquakes $N(m)$, 
having magnitude in the Richter scale greater than or equal to $m$, 
is phenomenologically observed to decrease with $m$ exponentially.
This gives the Gutenberg-Richter law
 \cite{bkc-Gutenberg:1954}
\[
{\rm ln} N(m) = {\rm constant} - a~ m,
\]

\noindent where $a$ is a constant. It appears \cite{bkc-Gutenberg:1954,bkc-Kostrov:1990}
that the amount of 
energy $\epsilon$ released in an earthquake of magnitude $m$
is related to it exponentially:
\[
{\rm ln} \epsilon = {\rm constant} + b~m,
\]

\noindent where $b$ is another constant. Combining therefore we get the
power law giving the number of earthquakes $N(\epsilon)$
releasing energy equal to $\epsilon$ as
\begin{equation}
\label{bkc-GR}
N(\epsilon) \sim \epsilon^{-\alpha},
\end{equation}

\noindent with $\alpha = a/b$. The observed value of the
exponent (power) $\alpha$ in (\ref{bkc-GR}) is around unity (see e.g., \cite{bkc-Kostrov:1990}).
 
Several laboratory and computer simulation models have 
recently been proposed \cite{bkc-Chakrabarti:1997} to capture essentially the above
power law in the earthquake energy release statistics. In a
very successful table-top laboratory simulation model of earthquakes, 
Burridge and Knopoff \cite{bkc-Burridge:1967} took a chain of wooden blocks connected by 
identical springs to the neighbouring blocks. The entire chain was
placed on a rigid horizontal table with a rough surface, and
one of the end blocks was pulled very slowly
and uniformly using a driving 
motor. The strains of the springs increase due to the creep motions 
of the blocks until one or a few of the blocks slip. The drops
in the elastic energy of the chain during slips could be 
measured from the extensions or compressions of all the springs, and
could be taken as the released energies in the earthquake. For some 
typical roughness of the surfaces (of the blocks and of the table),
the distribution of these drops in the elastic energy due to slips
indeed shows a power law behaviour with $\alpha \simeq$ 1 in (\ref{bkc-GR}). 

A computer simulation version of this model by Carlson and
Langer \cite{bkc-Carlson:1989} considers harmonic springs connecting equal mass
blocks which are also individually connected to a rigid frame
(to simulate other neighbouring portions of the earth's crust
not on the same tectonic plate) 
by harmonic springs. The entire
system moves on a rough surface with nonlinear velocity dependent 
force (decreasing to zero for large relative velocities) in the
direction opposite to  the relative motion between the block and
the surface. In the computer simulation of this model it is seen
that the distribution of the elastic energy release in such
a system can indeed be given by a power law like (\ref{bkc-GR}), provided the
nonlinearity of the friction force, responsible for the 
self-organisation, is carefully chosen \cite{bkc-Carlson:1989}. 

The lattice automata model of Bak et al \cite{bkc-Bak:1997}
represent the stress on each block by a height variable
at each lattice site. The site topples (the block
slips) if the height (or stress) at
that site exceeds a preassigned threshold value, and the height becomes 
zero  there and the neighbours share the stress by increasing their
heights by one unit. With this dynamics for the system,  if
any of the neighbouring sites  of the
toppled one was already at the threshold height,
the avalanche continues. The boundary sites are considered to be all
absorbing. With random addition of heights at a constant rate
(increasing stress at a constant rate due to 
tectonic motion), such
a system reaches its self-organised critical point where the
avalanche size distributions follow a natural power law corresponding
to this self-tuned critical state. Bak et al \cite{bkc-Bak:1997} identify 
this self-organised critical state to be responsible for the
Gutenberg-Richter type power law. 
All these models are successful in
capturing the Gutenberg-Richter power law, 
and the real reason
for the self-similarity inducing the power law is essentially the
same in all these different models: emergence of the self-oganised 
critical  state for wide yet suitably chosen variety of nonlinear
many-body coupled dynamics. In this sense all these models incoporate
the well-established fact of the stick-slip frictional instabilities
between the earth's crust and the tectonic plate. 
It is quite 
difficult to check at this stage any further details and predictions
of these models.

While the motion of the tectonic plate
 is surely an observed fact, 
and this stick-slip process should be a major ingredient of any
bonafide model of earthquake, another established fact regarding
the fault geometries of the earth's crust is the fractal nature 
of the roughness of the surfaces of the earth's crust and the
tectonic plate. This latter feature is
 missing in any of these models discussed
above.  In fact, the surfaces involved in the process are
results of large scale fracture seperating the crust from
the moving tectonic plate. Any such crack surface is observed
to be a self-similar fractal, having the 
self-affine scaling
property $z(\lambda x, \lambda y) \sim \lambda^{\zeta} z(x,y)$
for the surface coordinate $z$ in the direction perpendicular to 
the crack surface in the ($x,y$) plane \cite{bkc-Chakrabarti:1997}. 
Various fractographic
investigations indicate a fairly robust universal behaviour
for such surfaces and the roughness exponent 
$\zeta$ is observed to have
a value around $0.80-0.85$ (with a possible crossover to $\zeta
\simeq 0.4$ for slow propagation of the crack-tip)
\cite{bkc-Chakrabarti:1997,bkc-Sahimi:2003}. This widely
observed scaling property of the fracture surfaces also suggests that
the fault surfaces of the earth's crust or the 
tectonic plate 
should have similar fractal properties. In fact, some investigators 
of the earthquake dynamics have already pointed out that the fracture
mechanics of the stressed crust of the earth forms self-similar
fault patterns, with well-defined fractal dimensionalities near
the contact areas with the major plates \cite{bkc-Chakrabarti:1997}.
Based on these observations regarding the earthquake faults, we have
developed a `two-fractal overlap' model 
\cite{bkc-Chakrabarti:1999,bkc-Bhattacharyya:2005} discussed later.

\section{Fracture Statistics of Disordered Solids}

\subsection{Griffith Energy Balance and
 Brittle Fracture Strength of Solids}

One can easily show (see e.g. \cite{bkc-Lawn:1993,bkc-Chakrabarti:1997}) that
in a stressed solid, the local stress at sharp notches or
corners of the microcrack can rise to a level several times that of the applied
stress. This indicetes how the microscopic cracks or flaws within a solid
might become potential sources of weakness of the solid.
Although this stress concentration indicates clearly where the instabilities 
should occur, it is not sufficient to
tell us when the instability does occur and the 
fracture propagation starts.
This requires a detailed energy balance consideration.

Griffith in 1920, equating the released elastic energy (in an elastic contin-
uum) with the energy of the surface newly created (as the crack grows),
arrived at a quantitative criterion for the equilibrium extension of the 
microcrack already present within the stressed material
\cite{bkc-Bergman:1992}. We give below
an analysis which is valid effectively for two-dimensional stressed solids
with a single pre-existing crack, as for example the case of a large plate with
a small thickness. Extension to three-dimensional solids is straightforward.
\begin{figure}
\label{bkc-slitsigma}
\centering\resizebox*{4cm}{!}{\includegraphics{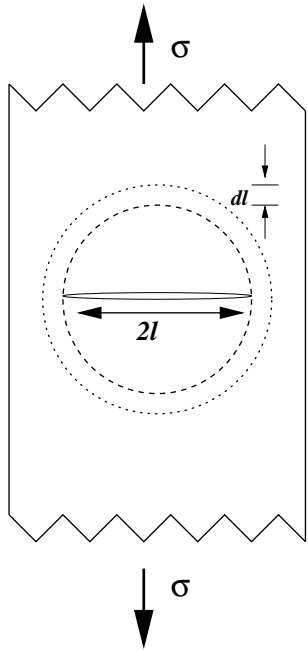}}
\caption{A portion of a plate (of thickness $w$) 
under tensile stress $\sigma$ (Model I loading)
containing a linear crack of length $2l$. For a further growth of the
crack length by $2{\rm d}l$, the elastic energy released from the
annular region must be sufficient to provide the surface energy 
$4 \Gamma w {\rm d} l$ (extra elastic energy must be released 
for finite velocity of crack propagation).}
\end{figure}

Let us assume a thin linear crack of length $2l$ in an infinite elastic
continuum subjected to uniform tensile stress 
$l$ perpendicular to the length
of the crack (see Fig. 1). Stress parallel to the crack does not affect the
stability of the crack and has not, therefore, been considered. Because of
the crack (which can not support any stress field, at least on its surfaces),
the strain energy density of the stress field ($\sigma^2 /2Y$) 
is perturbed in a region
around the crack, having dimension of the length of the crack. We assume
here this perturbed or stress-released region to have a circular cross-section
with the crack length as the diameter.
The exact geometry of this perturbed region is not important here, and it
determines only an (unimportant) numerical factor in the 
Griffith formula
(see e.g. \cite{bkc-Lawn:1993}).
Assuming therefore half of the stress energy of the annular or cylindrical
volume, having the internal radius $l$ and outer radius $l + {\rm d}l$ 
and length $w$ (perpendicular to the plane of the stress; here the width 
$w$ of the plate is very small compared to the other dimensions), 
to be released as the crack propagates by a length ${\rm d}l$, 
one requires this released strain energy to be sufficient for providing 
the surface energy of the four new surfaces produced. This suggests
\[
\frac{1}{2} (\sigma^2 /2Y) (2 \pi w l {\rm d}l) \ge \Gamma (4 w {\rm d}l).
\]
Here $Y$ represents the Young's modulus of the solid and $\Gamma$ represents the
surface energy density of the solid, measured by the extra energy required
to create unit surface area within the bulk of the solid.

We have assumed here, on average, half of the strain energy of the
cylindrical region having a circular cross-section with diameter $2l$ to be
released. If this fraction is different or the cross-section is different, 
it will
change only some of the numerical factors, in which we are not very much
interested here. Also, we assume here linear elasticity up to the breaking
point, as in the case of brittle materials. The equality holds when energy
dissipation, as in the case of plastic deformation or for the propagation
dynamics of the crack, does not occur. One then gets
\begin{equation}
\label{bkc-sigma_f}
\sigma_f = \frac{\Lambda}{\sqrt{2l}}; \ \ \Lambda = \left( \frac{4}{\sqrt \pi} \right) \sqrt{Y \Gamma}
\end{equation}
for the critical stress at and above which the crack of length $2l$ starts
propagating and a macroscopic fracture occurs. Here $\Lambda$ is called the 
stress-intensity factor or the fracture toughness. 
In fact, one can alternatively view the fracture occurring when the stress 
at the crack-tip (given by the stress intensity factor in (\ref{bkc-sigma_f}) 
exceeds the elastic stress limit for the medium.

In a three-dimensional solid containing a single elliptic disk-shaped pla-
nar crack perpendicular to the applied tensile stress direction, a straight-
forward extension of the above analysis suggests that the maximum stress
concentration would occur at the two tips (at the two ends of the major
axis) of the ellipse. The Griffith stress 
for the brittle fracture
 of the solid
would therefore be determined by the same formula (\ref{bkc-sigma_f}), 
with the crack length $2l$ replaced by the length of the major axis 
of the elliptic planar crack.
Generally, for any dimension therefore, if a crack of length $l$ already
exists in an infinite elastic continuum, subject to uniform tensile stress
$\sigma$ perpendicular to the length of the crack, then for the onset 
of brittle fracture , Griffith equates (the differentials of) the elastic 
energy $E_l$ with the surface energy $E_s$:
\begin{equation}
\label{bkc-E_l}
E_l \simeq \left( \frac{\sigma^2}{2Y}\right) l^d = E_s \simeq \Gamma l^{d-1},
\end{equation}
where $Y$ represents the elastic modulus appropriate for the strain, $\Gamma$
the surface energy density and $d$ the dimension. Equality holds when no energy
dissipation (due to plasticity or crack propagation) occurs and one gets
\begin{equation}
\label{bkc-sigma_f1}
\sigma_f \sim \frac{\Lambda}{\sqrt l}; \ \Lambda \sim \sqrt{Yl}
\end{equation}
for the breakdown stress at (and above) which the existing crack of length $l$
starts propagating and a macroscopic fracture occurs. It may also be noted
that the above formula is valid in all dimensions ($d \ge 2$).

For disordered solids, let us model the solid by a percolating system. As
mentioned earlier, for the occupied bond/site concentration $p > p_c$,
the percolation threshold, the typical pre-existing cracks in the solid
will have the dimension ($l$) of correlation length $\xi \sim \Delta p^{-\nu}$
and the elastic strength $Y \sim \Delta p^{T_e}$ \cite{bkc-Stauffer:1992}.
Assuming that the surface energy density $\Gamma$ scales as $\xi^{d_B}$, with
the backbone (fractal) dimension $d_B$ \cite{bkc-Stauffer:1992}, equating $E_l$
and $E_s$ as in (\ref{bkc-E_l}), one gets 
$\left( \frac{\sigma_f^2}{2Y}\right) \xi^d \sim \xi^{d_B}$.
This gives 
\[
\sigma_f \sim (\Delta p)^{T_f}
\]
with 
\begin{equation}
\label{bkc-T_f}
T_f = \frac{1}{2} [T_e + (d-d_B)\nu]
\end{equation}
for the `average' fracture strength of a disordered solid (of fixed value)
as one approaches the percolation threshold. Careful extensions of
such scaling relations (\ref{bkc-T_f}) and rigorous bounds for $T_f$ has been
obtained and compared extensively in 
\cite{bkc-Chakrabarti:1997,bkc-Sahimi:2003}.

\subsection{Extreme Statistics of The Fracture Stress}

The fracture strength $\sigma_f$ of a disordered solid does not have
self-averaging statistics; most probable and the average $\sigma_f$
may not match because of the extreme
 nature of the statistics. This is
because, the `weakest point' of a solid determines the strength of the
entire solid, not the average weak points!
As we have modelled here, the statistics of clusters of defects are
governed by the random percolation processes. We have also
discussed, how the linear responses, like the elastic moduli
of such random networks, can be obtained from the averages
over the statistics of such clusters. This was possible because of the 
self-averaging property of such linear responses.
This is because the elasticity of a random network is determined by
all the `parallel' connected material portions or paths, 
contributing their share in the net
elasticity of the sample. However, the fracture or breakdown property of a 
disordered solid is determined by only the weakest (often the longest) defect
cluster or crack in the entire solid. Except for some indirect effects, most
of the weaker or smaller defects or cracks in the solid do not determine the
breakdown strength of the sample. The fracture 
or breakdown statistics of
a solid sample is therefore determined essentially by the 
extreme statistics
of the most dangerous or weakest (largest) defect cluster or crack within
the sample volume. 

We discuss now more formally the origin of this 
extreme statistics.
Let us consider a solid of linear size $L$, containing $n$ cracks within
its volume. We assume that each of these cracks have a failure probability
$f_i(\sigma), i=1,2,\ldots,n$ to fail or break (independently) under an
applied stress $\sigma$ on the solid, and that the perturbed or
stress-released regions of each of these cracks are seperate and do not 
overlap. If we denote the cumulative failure probability of the entire
sample, under stress $\sigma$, by $F(\sigma)$ then \cite{bkc-Chakrabarti:1997}
\begin{equation}
\label{bkc-1-Fsigma}
1-F(\sigma)=\prod_{i=1}^{n} (1-f_i(\sigma)) \simeq 
\exp \left[ -\sum_i f_i(\sigma) \right] = \exp \left[ -L^d \tilde{g}(\sigma) \right]
\end{equation}
where $\tilde{g}(\sigma)$ denotes the density of cracks within the sample
volume $L^d$ (coming from the sum $\sum_i$ over the entire volume),
which starts propagating at and above the stress level $\sigma$. The above
equation comes from the fact that the sample survives if each of the cracks
within the volume survives. This is the essential origin of the above
extreme statistical
 nature of the failure probability $F(\sigma)$ of the
sample.

Noting that the pair correlation $g(l)$ of two occupied sites at distance $l$
on a percolation cluster decays as $\exp\left(-l/\xi(p)\right)$, and
connecting the stress $\sigma$ with the length $l$ by using 
Griffith's law
(\ref{bkc-sigma_f}) that $\sigma \sim \frac{\Lambda}{l^\phi}$, one gets
$\tilde{g}(\sigma) \sim \exp \left( -\frac{\Lambda^{1/\phi}}{\xi \sigma^{1/\phi}} \right)$
for $p \to p_c$. This, put in eqn. (\ref{bkc-1-Fsigma}) gives the Gumbel distribution 
(\ref{bkc-gumbell}) given earlier \cite{bkc-Chakrabarti:1997}. If, on the other
hand, one assumes a power law decay of $g(l)$: $g(l) \sim \l^{-w}$, then
using the Griffith's law (\ref{bkc-sigma_f}), one gets 
$\tilde{g}(\sigma) \sim \left( \frac{\sigma}{\Lambda} \right)^m$, giving
the Weibull distribution (\ref{bkc-weibull}), 
from eqn. (\ref{bkc-1-Fsigma}),
where $m=w/\phi$ gives the Weibull modulus 
\cite{bkc-Chakrabarti:1997}.
The variation of $F(\sigma)$ with $\sigma$ in both the cases
have the generic form shown in Fig. 2. $F(\sigma)$ is non-zero for any 
stress $\sigma > 0$ and its value (at any $\sigma$) is higher for larger 
volume ($L^d$).
This is because, the possibility of a larger defect (due to fluctuation)
is higher in a larger volume and consequently, its failure probability
is higher. Assuming $F(\sigma_f)$ is finite for failure, the most probable
failure stress $\sigma_f$ becomes a decreasing function of volume if
extreme statistics at work.
\begin{figure}
\label{bkc-weibullcurve}
\centering\resizebox*{8cm}{!}{\includegraphics{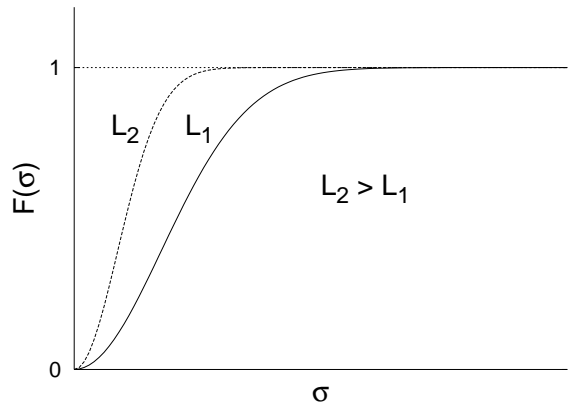}}
\caption{Schematic variation of failure probability $F(\sigma)$ with stress 
$\sigma$ for a disordered solid with volume $L_1^d$ or $L_2^d$ ($L_2 > L_1$).}
\end{figure}

The precise ranges of the validity of the Weibull or 
Gumbel 
distributions for the breakdown strength of disordered solids are not 
well established yet. However, analysis of the results of detailed 
experimental and numerical studies of breakdown in disordered solids seem 
to suggest that the fluctuations of the extreme statistics
 dominate for 
small disorder \cite{bkc-Sahimi:2003}. Very near to the percolation point,
the percolation statistics takes over and the 
statistics
become self-averaging. One can argue \cite{bkc-Bergman:1992}, that arbitrarily 
close to the percolation threshold, the 
fluctuations
of the extreme statistics
 will probably get suppressed and the percolation 
statistics should take over and the most probable breaking stress 
becomes independent of the sample volume (its variation with 
disorder being determined, as in Eqn.(\ref{bkc-T_f}), by an 
appropriate breakdown exponent). This is because the appropriate 
competing length scales for the two kinds of statistics are the
Lifshitz scale $\ln L$ (coming from the finiteness of the volume integral of the
defect probability: $L^d(1 - p)^l$ finite, giving the typical defect size 
$l \sim \ln L$) and the percolation 
correlation length $\xi$. 
When $\xi < \ln L$, the above scenario of extreme statistics should be 
observed. For $\xi > \ln L$, the percolation statistics is expected 
to dominate. 

\subsection{Failure Statistics in Fiber Bundles}
 
\begin{figure}
\resizebox{7.0cm}{!}{\includegraphics{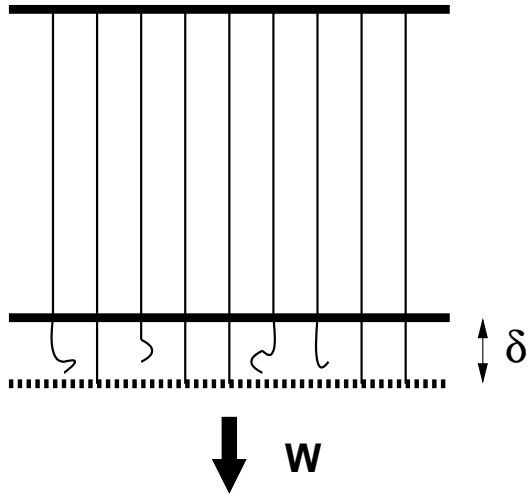}}
\caption{
The fiber bundle consists initially of $N$ fibers attached in parallel
to a fixed and rigid plate at the top and a downwardly movable platform 
from which a load $W$ is suspended at the bottom. In the 
equal load sharing
model considered here, the platform is absolutely rigid and the load $W$ is
consequently shared equally by all the intact fibers.}
\label{bkc-fbm-schematic}
\end{figure}

The fiber bundle (see Fig. \ref{bkc-fbm-schematic}) consists of $N$ fibers 
or Hook springs, each having
identical spring constant $\kappa$. The bundle supports a load 
$W=N\sigma$ and the breaking threshold $\left( \sigma _{th}\right) _{i}$
of the fibers are assumed to be different for different fiber ($i$).
For the equal load sharing model we consider here, the lower
platform is absolutely rigid, and therefore no local deformation and hence 
no stress concentration occurs anywhere around the failed fibers. This
ensures equal load sharing, i.e., the intact fibers share
the applied load $W$ equally and the load per fiber increases as
more and more fibers fail. The strength of 
each of the fiber $\left( \sigma_{th}\right)_{i}$ in the bundle is 
given by the stress value it can bear, and beyond which
it fails. The strength of the fibers are taken from a randomly distributed
normalised density $\rho (\sigma _{th})$ within the interval
$0$ and $1$ such that 
\[
\int _{0}^{1}\rho (\sigma _{th})d\sigma _{th}=1.
\] 
The equal load sharing assumption neglects `local' fluctuations
in stress (and its redistribution) and renders the model as a mean-field
one. 

The breaking dynamics starts when an initial stress \( \sigma  \)
(load per fiber) is applied on the bundle. The fibers having strength
less than \( \sigma  \) fail instantly. Due to this rupture, total
number of intact fibers decreases and rest of the (intact) fibers
have to bear the applied load on the bundle. Hence effective stress
on the fibers increases and this compels some more fibers to break.
These two sequential operations, namely the stress redistribution and further
breaking of fibers continue till an equilibrium is reached, where
either the surviving fibers are strong enough to bear the applied
load on the bundle or all fibers fail.

\begin{figure}
\label{bkc-uniform}
\centering\resizebox*{7cm}{!}{\includegraphics{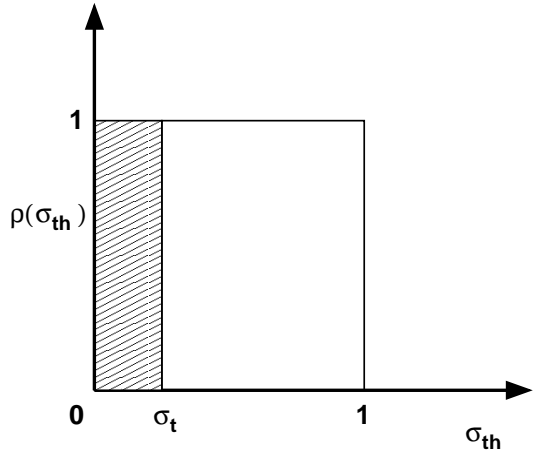}}
\caption{
The simple model considered here assumes uniform density $\rho (\sigma _{th})$
of the fiber strength distribution up to a cutoff strength (normalized to 
unity). At any load per fiber level $\sigma_t$ at time $t$, the fraction 
$\sigma_t$ fails and $1-\sigma_t$ survives.
}
\end{figure}

This breaking dynamics can be represented by recursion
relations in discrete time steps. 
For this, let us consider a very simple model of fiber bundles where
the fibers (having the same spring constant $\kappa$) have a white
or uniform strength distribution  $\rho(\sigma_{th})$ upto a cutoff strength
normalized to unity, as shown in Fig. \ref{bkc-uniform}:
$\rho (\sigma_{th}) = 1$ for $0 \le \sigma_{th} \le 1$ and 
$= \rho(\sigma_{th})=0$ for $\sigma > \sigma_{th}$.
Let us also define $U_t(\sigma)$ to be the fraction of fibers in the bundle
that survive after (discrete) time step $t$, counted from the time $t=0$
when the load is put (time step indicates the number of stress 
redistributions). As such, $U_t(\sigma=0)=1$ for all $t$ and $U_t(\sigma)=1$
for $t=0$ for any $\sigma$; 
$U_t(\sigma)=U^*(\sigma) \ne 0$ for $t \to \infty$ and 
$\sigma < \sigma_c$, the critical or failure strength of the bundle, and 
$U_t(\sigma)=0$ for $t \to \infty$ if $\sigma > \sigma_c$.

Therefore $U_{t}(\sigma)$ follows a simple recursion relation 
(see Fig. \ref{bkc-uniform})
\[
U_{t+1}= 1-\sigma_t;\ \ \sigma_t = \frac{W}{U_t N}
\]
\begin{equation}
\label{bkc-recU_t}
{\rm or,} \ \ U_{t+1}=1-\frac{\sigma }{U_{t}}.
\end{equation}
At the equilibrium state (\( U_{t+1}=U_{t}=U^{*} \)), the above relation
takes a quadratic form of \( U^{*} \) : 
\[
U^{*^{2}}-U^{*}+\sigma =0.
\]

\noindent The solution is

\[
U^{*}(\sigma )=\frac{1}{2}\pm (\sigma_{f}-\sigma )^{1/2};\sigma_{f}=\frac{1}{4}.
\]

\noindent Here \( \sigma_{f} \) is the critical value of initial
applied stress beyond which the bundle fails completely. The solution
with (\( + \)) sign is the stable one, whereas the one with (\( -) \)
sign gives unstable solution \cite{bkc-Pradhan:2001}. The quantity
\( U^{*}(\sigma ) \) must be real valued as it has a physical meaning:
it is the fraction of the original bundle that remains intact under
a fixed applied stress \( \sigma  \) when the applied stress lies
in the range \( 0\leq \sigma \leq \sigma_{f} \). Clearly, \( U^{*}(0)=1 \).
Therefore the stable solution can be written as 
\begin{equation}
\label{bkc-Ustarsigma_c}
U^{*}(\sigma )=U^{*}(\sigma_{f})+(\sigma_{f}-\sigma )^{1/2};
\ U^*(\sigma_f) = \frac{1}{2} \ {\rm and}\ \sigma_{f}=\frac{1}{4}.
\end{equation}
For \( \sigma >\sigma_{f} \) we can not get a real-valued fixed
point as the dynamics never stops until \( U_{t}=0 \) when the bundle
breaks completely.

\vskip.2in
\noindent \textbf{(a) At \(\sigma <\sigma_{f} \)}
\vskip.2in
It may be noted that the quantity \( U^{*}(\sigma )-U^{*}(\sigma_{f}) \)
behaves like an order parameter that determines a transition from
a state of partial failure (\( \sigma \leq \sigma_{f} \)) to a state
of total failure (\( \sigma >\sigma_{f} \)) \cite{bkc-Pradhan:2001}:
\begin{equation}
\label{bkc-Ustar}
O\equiv U^{*}(\sigma )-U^{*}(\sigma_{f})=(\sigma_{f}-\sigma )^{\beta };\beta =\frac{1}{2}.
\end{equation}

To study the dynamics away from criticality ($\sigma \rightarrow \sigma_{f}$
from below), we replace the recursion relation (\ref{bkc-recU_t}) by a differential
equation 
\[
-\frac{dU}{dt}=\frac{U^{2}-U+\sigma }{U}.
\]

\noindent Close to the fixed point we write \( U_{t}(\sigma )=U^{*}(\sigma ) \)
+ \( \epsilon  \) (where \( \epsilon \rightarrow 0 \)). This, following
Eq. (10), gives \cite{bkc-Pradhan:2001} 
\begin{equation}
\label{bkc-epsilon}
\epsilon =U_{t}(\sigma )-U^{*}(\sigma )\approx \exp (-t/\tau ),
\end{equation}

\noindent where \( \tau =\frac{1}{2}\left[ \frac{1}{2}(\sigma_{f}-\sigma )^{-1/2}+1\right]  \).
Near the critical point we can write \begin{equation}
\label{bkc-dec19}
\tau \propto (\sigma_{f}-\sigma )^{-\alpha };\alpha =\frac{1}{2}.
\end{equation}
 Therefore the relaxation time diverges following a power-law as \( \sigma \rightarrow \sigma_{f} \)
from below \cite{bkc-Pradhan:2001}.

One can also consider the breakdown susceptibility \( \chi  \), defined
as the change of \( U^{*}(\sigma ) \) due to an infinitesimal increment
of the applied stress \( \sigma  \) \cite{bkc-Pradhan:2001} \begin{equation}
\label{bkc-sawq}
\chi =\left| \frac{dU^{*}(\sigma )}{d\sigma }\right| =\frac{1}{2}(\sigma_{f}-\sigma )^{-\gamma };\gamma =\frac{1}{2}
\end{equation}

\noindent from equation (10). Hence the susceptibility diverges as
the applied stress \( \sigma  \) approaches the critical value \( \sigma_{f}=\frac{1}{4} \).
Such a divergence in \( \chi  \) had already been observed in the
numerical studies.

\vskip.2in
\noindent \textbf{(b) At} \textbf{\large \(\sigma =\sigma_{f} \)}{\large\par}
\vskip.2in
\noindent At the critical point (\( \sigma =\sigma_{f} \)), we observe
a different dynamic critical behavior in the relaxation of the failure process.
From the recursion relation (\ref{bkc-recU_t}), it can be shown
that decay of the fraction \( U_{t}(\sigma_{f}) \) of unbroken fibers
that remain intact at time \( t \) follows a simple power-law decay
\cite{bkc-Pradhan:2001}:
\begin{equation}
\label{bkc-qqq}
U_{t}=\frac{1}{2}(1+\frac{1}{t+1}),
\end{equation}

\noindent starting from \( U_{0}=1 \). For large \( t \) (\( t\rightarrow \infty  \)),
this reduces to \( U_{t}-1/2\propto t^{-\delta } \); \( \delta =1 \);
a strict power law which is a robust characterization of the critical
state.

\subsubsection{Universality Class of The Model}

The universality class of the model has been checked \cite{bkc-Pradhan:2001} taking
two other types of fiber strength distributions: (I) linearly increasing
density distribution and (II) linearly decreasing density distribution
within the ($\sigma_{th}$) limit $0$ and $1$. One can show that while 
$\sigma_{f}$ changes with different strength distributions 
($\sigma_f= \sqrt{4/27}$ for case (I) and $\sigma_f=4/27$ for case II), 
the critical behavior remains unchanged: $\alpha =1/2=\beta =\gamma$, 
$\delta =1$ for all these equal load sharing models.

\subsubsection{Nonlinear Stress-Strain Relation for The Bundle}

\noindent One can now consider a slightly modified strength distribution
of the equal load sharing fiber bundle, showing typical nonlinear deformation 
characteristics \cite{bkc-Daniels:1945,bkc-Pradhan:2001}. For this, we consider 
an uniform density
distribution of fiber strength, having a lower cutoff. Until failure
of any of the fibers (due to this lower cutoff), the bundle shows
linear elastic behavior. As soon as the fibers start failing, the
stress-strain relationship becomes nonlinear. The dynamic critical
behavior remains essentially the same and the static (fixed point)
behavior shows elastic-plastic like deformation before rupture of the bundle.

Here the fibers are elastic in nature having identical force constant
\( \kappa  \) and the random fiber strengths distributed uniformly
in the interval \( [\sigma _{L},1] \) with \( \sigma _{L}>0 \);
the normalised distribution of the threshold stress of the fibers
thus has the form (see Fig. 5): \begin{equation}
\label{bkc-jan31-mod}
\rho (\sigma _{th})=\left\{ \begin{array}{cc}
0, & 0\leq \sigma _{th}\leq \sigma _{L}\\
\frac{1}{1-\sigma _{L}}, & \sigma _{L}<\sigma _{th}\leq 1
\end{array}\right. .
\end{equation}

\begin{figure}
\label{bkc-fig:1-sigmaL} 
\centering\resizebox*{6cm}{!}{\includegraphics{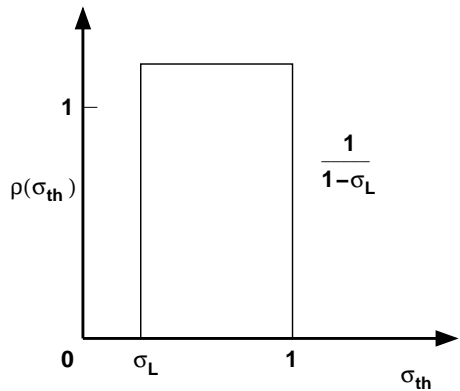}}
\caption{
The fiber breaking strength distribution \( \rho (\sigma _{th}) \)
considered for studying elastic-plastic type nonlinear deformation
behavior of the equal load sharing model.
}
\end{figure}

For an applied stress \( \sigma \leq \sigma _{L} \) none of the fibers
break, though they are elongated by an amount \( \delta  \) \( =\sigma /\kappa  \).
The dynamics of breaking starts when applied stress \( \sigma  \)
becomes greater than \( \sigma _{L} \). Now, for \( \sigma >\sigma _{L} \)
the fraction of unbroken fibers follows a recursion relation (for
\( \rho (\sigma _{th}) \) as in Fig. 5): \begin{equation}
\label{bkc-abcdef}
U_{t+1}=1-\left[ \frac{F}{NU_{t}}-\sigma _{L}\right] \frac{1}{1-\sigma _{L}}=\frac{1}{1-\sigma _{L}}\left[ 1-\frac{\sigma }{U_{t}}\right] ,
\end{equation}

\noindent which has stable fixed points: \begin{equation}
\label{bkc-feb7c}
U^{*}(\sigma )=\frac{1}{2(1-\sigma _{L})}\left[ 1+\left( 1-\frac{\sigma }{\sigma _{f}}\right) ^{1/2}\right] ;\sigma _{f}=\frac{1}{4(1-\sigma _{L})}.
\end{equation}
\noindent The model now has a critical point $\sigma _{f}=1/[4(1-\sigma _{L})]$
beyond which total failure of the bundle takes place. The above equation
also requires that \( \sigma _{L}\leq 1/2 \) (to keep the fraction
\( U^{*}\leq 1 \)). As one can easily see, the dynamics of \( U_{t} \)
for \( \sigma <\sigma _{f} \) and also at \( \sigma =\sigma _{f} \)
remains the same as discussed in the earlier section. At each fixed
point there will be an equilibrium elongation \( \delta (\sigma ) \)
and a corresponding stress \( S=U^{*}\kappa \delta (\sigma ) \) develops
in the system (bundle). This \( \delta (\sigma ) \) can be easily
expressed in terms of \( U^{*}(\sigma ) \). This requires the evaluation
of \( \sigma ^{*} \), the internal stress per fiber developed at
the fixed point, corresponding to the initial (external) stress \( \sigma  \)
(\( =F/N \)) per fiber applied on the bundle when all the fibers
were intact. Expressing the effective stress $\sigma^*$ per fiber in terms of
$U^*(\sigma)$, one can write from (\ref{bkc-abcdef})
\[
U^{*}(\sigma )=1-\frac{\sigma ^{*}-\sigma _{L}}{(1-\sigma _{L})}=\frac{1-\sigma ^{*}}{1-\sigma _{L}},
\]
for $\sigma > \sigma_L$. Consequently, 
\[
\kappa \delta (\sigma )=\sigma ^{*}=1-(1-\sigma _{L})U^{*}(\sigma ).
\]
 It may be noted that the internal stress 
$\sigma _{c}^{*}$ ($ =\sigma _{c}/U^{*}(\sigma _{c})$ is universally 
equal to $1/2$ (independent of $\sigma_{L}$; from (\ref{bkc-feb7c})) 
at the failure point \( \sigma =\sigma _{c} \) of the
bundle. This finally gives the stress-strain relation for the
model : \begin{equation}
\label{bkc-may22}
S=\left\{ \begin{array}{cc}
\kappa \delta , & 0\leq \sigma \leq \sigma _{L}\\
\kappa \delta (1-\kappa \delta )/(1-\sigma _{L}), & \sigma _{L}\leq \sigma \leq \sigma _{f}\\
0, & \sigma >\sigma _{f}
\end{array}\right. .
\end{equation}

\begin{figure}
\label{bkc-stress-strain}
\centering\resizebox*{8cm}{!}{\includegraphics{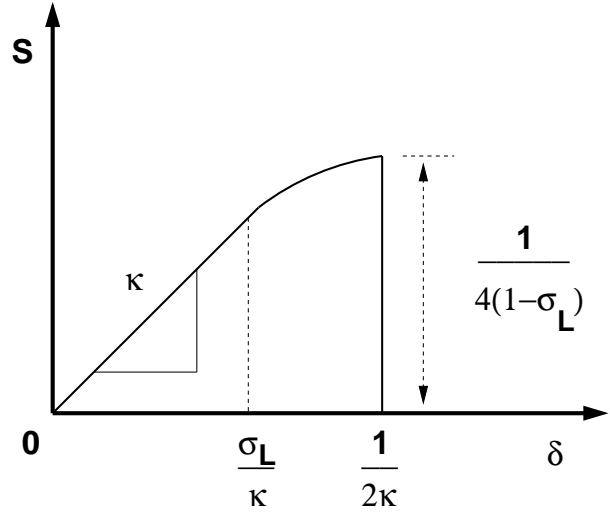}}
\caption{
 Schematic
stress (\( S \))-strain (\( \delta  \)) curve of the bundle (shown
by the solid line), following Eq. (\ref{bkc-may22}), 
with the fiber strength distribution (\ref{bkc-jan31-mod}) (as shown in Fig. 5).
Note, the model gives analytic solution for the full nonlinear
stress-strain relationship of the bundle (disordered solid), including
its failure (fracture) stress  or strain.
}
\end{figure}

This stress-strain relation is schematically shown in Fig.
6, where the initial linear region has slope \( \kappa  \) (the force
constant of each fiber). This Hooke's region for stress \( S \) continues
up to the strain value \( \delta =\sigma _{L}/\kappa  \), until which
no fibers break (\( U^{*}(\sigma )=1 \)). After this, nonlinearity
appears due to the failure of a few of the fibers and the consequent
decrease of \( U^{*}(\sigma ) \) (from unity). It finally drops to
zero discontinuously by an amount \( \sigma _{f}^{*}U^{*}(\sigma _{f})=1/[4(1-\sigma _{L})]=\sigma _{f} \)
at the breaking point \( \sigma =\sigma _{f} \) or \( \delta =\sigma ^{*}_{f}/\kappa =1/2\kappa  \)
for the bundle. This indicates that the stress drop at the final failure
point of the bundle is related to the extent (\( \sigma _{L} \))
of the linear region of the stress-strain curve of the same bundle.

\subsubsection{Strength of The 
Local Load Sharing Fiber Bundles}

So far, we studied models with fibers sharing the external load equally.
This type of model shows (both analytically and
numerically) existence of a critical strength (non zero $\sigma_{f}$)
of the macroscopic bundle \cite{bkc-Pradhan:2001} beyond which it collapses. 
The other extreme model, i.e., the local load sharing
model has been proved to be difficult to tackle analytically.

It is clear, however, that the extreme statistics
 comes into play for such
loacl load sharing models, for which the strength $\sigma_f \to 0$ as the
bundle size ($N$) approaches infinity. Essentially, for any finite load
($\sigma$), depending on the fiber strength distribution, the size of the
defect cluster can be estimated using 
Lifshitz argument (see section 2.2)
as $\ln N$, giving the failure strength $\sigma_f \sim 1/(\ln N)^a$,
where the exponent $a$ assumes a value appropriate for the model 
(see e.g., \cite{bkc-Pradhan:2003}).
If a fraction $f$ of the load of the failed fiber goes for global
redistribution and the rest (fraction $1-f$) goes to the fibers
neighboring to the failed one, then we see (see Pradhan et al 
\cite{bkc-Pradhan:2001}) that there is a crossover from extreme to
self-averaging statistics at a finite value of $f$.

\section{Two Fractal Overlap Model of Earthquake and its Statistics}

Overlapping fractals form a whole class of models to simulate
earthquake dynamics. These models are motivated by the
observation that a fault surface, like a fractured surface,
is a fractal object \cite{bkc-Chakrabarti:1997,bkc-Sahimi:2003}.
Consequently a fault may be viewed as a pair of overlapping fractals.
Fractional Brownian profiles have been commonly used as models of fault
surfaces \cite{bkc-Sahimi:2003}.
In that case the dynamics of a fault is represented by one Brownian
profile drifting on another and each intersection of the two profiles
corresponds to an earthquake \cite{bkc-Rubeis:1996}. However
the simplest possible model of a fault $-$ from the fractal point of
view $-$ was proposed by Chakrabarti and Stinchcombe \cite{bkc-Chakrabarti:1999}.
This model is a schematic representation of a fault by a pair of dynamically
overlapping Cantor sets. It is not realistic but, as a system of overlapping
fractals, it has the essential feature. Since the Cantor set is a fractal
with a simple construction procedure, it allows us to study in detail the
statistics of the overlap of one fractal object on another.
The two fractal overlap magnitude changes in time as one fractal moves
over the other. The overlap (magnitude)
time series can therefore be studied as a model time series of earthquake
avalanche dynamics \cite{bkc-Carlson:1989}.
\begin{figure}
\centering\resizebox*{14cm}{!}{\includegraphics{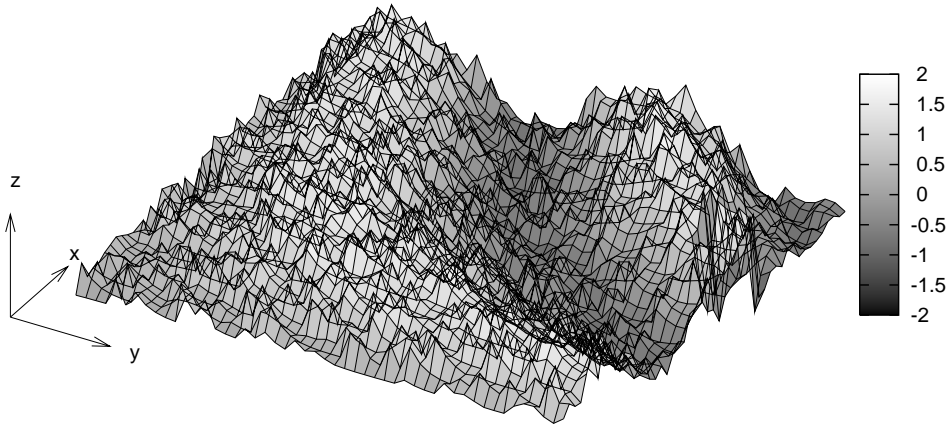}}
\caption{
A typical fracture surface: it has the 
self-affine scaling property 
$z(\lambda x, \lambda y)= \lambda^\zeta z(x,y)$ where the roughness
exponent $\zeta$ has some universal value (e.g., $\zeta \sim 0.8$ for
$(2+1)$-dimensional fractured surface).
}
\label{bkc-selfaffine}
\end{figure}

The statistics of overlaps between two fractals is
not studied much yet, though their knowledge is often required in various
physical contexts. It has been established recently that since
the fractured surfaces have got well-characterized self-affine properties,
the distribution of the elastic energies
released during the slips between two fractal surfaces (earthquake
events) may follow the overlap distribution of two self-similar fractal
surfaces \cite{bkc-Chakrabarti:1999,bkc-Pradhan:2003}.
Chakrabarti and Stinchcombe \cite{bkc-Chakrabarti:1999}
had shown analytically by renormalization group calculations that for
regular fractal overlap (Cantor sets 
and carpets) the contact area
distribution $\rho(s)$ follows a simple power law decay:
\begin{equation}
\rho (s) \sim s^{- \gamma}; \; \gamma = 1.
\label{bkc-eq:GR}
\end{equation}

In this so called 
Chakrabarti-Stinchcombe model 
\cite{bkc-Bhattacharyya:2005}, 
the solid-solid  contact surfaces of both the earth's
crust and the tectonic plate are considered as average self-similar
fractal surfaces. We then consider the distribution of contact
areas, as one fractal surface slides over the other.
We relate the total contact area between the two surfaces to
be proportional to the elastic strain energy that can be grown
during the sticking period, as the solid-solid friction force
arises from the elastic strains at the contacts between the asperities.
 We then consider this energy to be released as one surface
slips over the other and sticks again to the next contact or
overlap between the rough surfaces.
Since the two fractured surfaces are well known 
fractals, with
established (self-affine) scaling properties (see Fig. \ref{bkc-selfaffine})
Considering that such slips occur at intervals  proportional
to the length corresponding to that area, we obtain a power
law for the frequency distribution of the energy releases. This 
compares quite well with the 
Gutenberg-Richter law.

\begin{figure}
\centering\resizebox*{10cm}{!}{\includegraphics{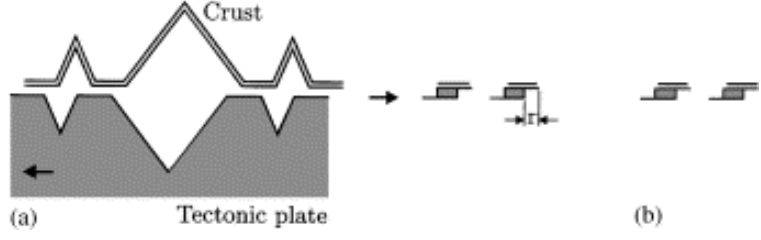}}
\caption{
Schematic representations of a portion of the
rough surfaces of the earth's crust and the supporting (moving)
tectonic plate. (b) The one dimensional projection of the surfaces
form Cantor sets of varying contacts or overlaps as one surface
slides over the other.}
\label{bkc-prfig1}
\end{figure}

\begin{figure}
\centering\resizebox*{12cm}{!}{\includegraphics{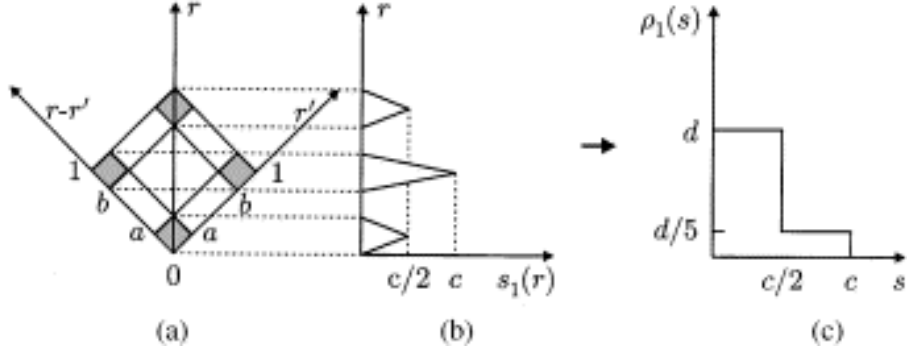}}
\caption{
(a) Two cantor sets (in their
first generation) along the axes $r$ and $r-r'$. (b) This gives
the overlap $s_1(r)$ along the diagonal. (c) The corresponding
density $\rho_1(s)$ of the overlap $s$ at this generation.
}
\label{bkc-prfig2}
\end{figure}

In order to proceed with the estimate of the number density 
$n(\epsilon)$ of earthquakes releasing energy $\epsilon$ 
in our model, we first find out the distribution  
$\rho (s)$ of the overlap or contact area $s$ between two 
self-similar fractal surfaces. We then relate $s$ with $\epsilon$
and the frequency of slips as a function of $s$, giving finally
 $n(\epsilon)$. To start with a simple problem of contact
area distribution between two fractals, we first take two Cantor
sets \cite{bkc-Chakrabarti:1999}  to 
model the contact area variations of two (nonrandom and self-similar)
surfaces as one surface  slides
over the other. Figure \ref{bkc-prfig1}(a) depicts structure in such surfaces
at a scale which corresponds to only the second generation of
iterative construction of two displaced Cantor sets, shown 
in Fig. \ref{bkc-prfig2}(b). It is obvious that with successive iterations, 
these surfaces will acquire self-similarity at every length
scale, when the generation number goes to infinity. We intend to
study the distribution of the total overlap $s$ (shown by the
shaded regions in Fig. \ref{bkc-prfig2}(b)) between the two Cantor sets, in
the infinite generation limit. 

\begin{figure}
\centering\resizebox*{12cm}{!}{\includegraphics{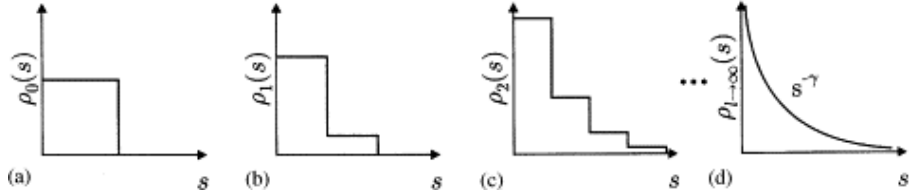}}
\caption{
The overlap densities $\rho (s)$ at various
generations of the Cantor sets
 at the zeroth (a), first (b),
second (c) and at the infinite (or fixed point) (d) generations.
}
\label{bkc-prfig3}
\end{figure}

Let the sequence of generators 
$G_l$ define our Cantor sets within the interval [0,1]: $G_0
= [0,1], G_1  \equiv RG_0 = [0,a] \bigcup
[b,1] $ (i.e., the union of the intervals $[0,a]$ and $[b,1]),
~ ...~, G_{l+1} = RG_l, ~ ... ~$. If we represent the
mass density of the set $G_l$ by $D_l(r)$, then $D_l(r)$ = 1 if $r$
is in any of the occupied intervals of $G_l$, and $D_l(r)$ = 0 
elsewhere. The required overlap magnitude between the sets
at any generation $l$ is then given by the 
convolution form  $s_l(r) = \int dr'
D_l(r') D_l(r-r')$. This form applies to symmetric fractals (with
$D_l(r) = D_l(-r)$); in general the argument of the second $D_l$
should be $D_l(r+r')$. 

One can express the overlap integral $s_1$ in the first generation by 
the projection  of the shaded regions along the 
vertical diagonal in Fig. 8(a). That gives the form 
shown in Fig. 8(b).
 For $a=b \le {1\over 3} $, the
nonvanishing $s_1(r)$ regions do not overlap, and 
are symmetric on both sides with the slope of the middle
curve being exactly double  those on the sides. One can then
easily check that the distribution $\rho_1(s) $ of  overlap $s$ at this
generation is given by Fig. 8(c), with both $c$ and $d$ greater
than unity, maintaining the normalisation of the probability
$\rho_1$ with $cd = 5/3$. The successive generations of the 
density $\rho_l(s)$ may thefore be represented by Fig. \ref{bkc-prfig2},
where
\begin{equation}
\label{bkc-rho_l+1}
\rho_{l+1}(s) = \tilde R \rho_l(s) \equiv {d\over 5}
\rho_l \left({s\over c}\right) + {4d\over 5}\rho_l \left(
{2s\over c}\right).
\end{equation}
In the infinite generation limit of the
renormalisation group (RG) equation, if $\rho^*(s) $ denotes
the fixed point distribution such that $\rho^*(s) = \tilde R
\rho^*(s)$, then assuming $\rho^*(s) \sim s^{-\gamma} \tilde
\rho(s)$, one gets $(d/5)c^{\gamma} + (4d/5)(c/2)^{\gamma} =$
1. Here $\tilde \rho(s)$ represents an arbitrary modular
function, which also
includes a logarithmic correction for large $s$. This
agrees with the above mentioned normalisation condition 
$cd = 5/3$ for the choice $\gamma = 1$.   This result 
for the overlap distribution (\ref{bkc-eq:GR})
\[
\rho^*(s) \equiv \rho(s) \sim s^{-\gamma}; ~~ \gamma =1,
\]
is the general result for all cases that
we have investigated and solved by the functional rescaling
technique (with the $\log s$ correction for large $s$,
renormalising the total integrated distribution).
\begin{figure}
\centering\resizebox*{10cm}{!}{\includegraphics{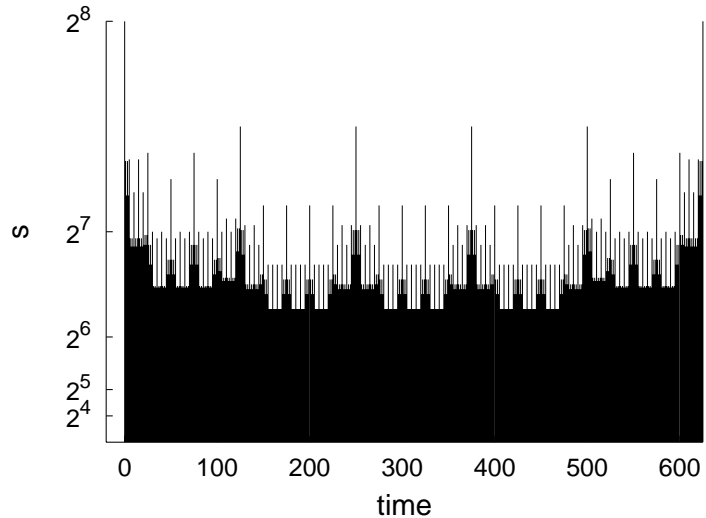}}
\caption{
For two (regular) Cantor sets, one moving uniformly over the other 
(with periodic boundary condition), the total measure of the (shaded region
in Fig. \ref{bkc-prfig1}(b) contributing to the) overlap $s$ has the time variation
as shown here (for $n=4$).
}
\label{bkc-timeseries}
\end{figure}

The above study is for continuous relative motion of one
Cantor set over the other.
Study of the time ($t$) variation of contact area (overlap)
$s(t)$ between two well-characterized fractals having the same
fractal dimension as one fractal moves 
over the other with constant velocity, 
with discrete (minimum element in that generation) steps 
has been studied, for finite generations \cite{bkc-Bhattacharyya:2005}. 
Bhattacharyya \cite{bkc-Bhattacharyya:2005} studied this overlap
distribution for two Cantor sets with periodic boundary conditions and
each having dimension $\log 2/\log 3$ (see Fig. \ref{bkc-timeseries}).
It was shown, using exact counting, that if $s \equiv 2^{n-k}$ ($n$ is the
generation number) then the probability $\tilde{\rho}(s)$ to get an overlap $s$
is given by a binomial distribution \cite{bkc-Bhattacharyya:2005}
\begin{equation}
\label{bkc-binomial}
\tilde{\rho}(2^{n-k}) = \left ( \begin{array}{c} n\\ n-k \end{array} \right )
 \left ( {1 \over 3} \right )^{n-k} \left ( {2 \over 3} \right )^k
\; \sim \exp(-r^2/n); \; r \to 0,
\end{equation}
where $r^2 = \left[ \frac{3}{2} \left( \frac{2}{3} n - k\right)\right]^2$.
Expressing therefore $r$ by $\log s$ near the maxima of $\tilde{\rho}(s)$, one
can again rewrite (\ref{bkc-binomial}) as
\begin{equation}
\label{bkc-dtildes}
\tilde{\rho}(s) \sim \exp \left( - \frac{(\log s)^2}{n} \right); 
\; n \to \infty.
\end{equation}
Noting that $\tilde{\rho}(s) {\rm d}(\log s) \sim \rho(s) {\rm d} s$, we find
$\rho(s) \sim s^{-\gamma}$, $\gamma=1$, as in (\ref{bkc-eq:GR}),
as the binomial or Gaussian part becomes a very weak function of $s$
as $n \to \infty$ \cite{bkc-Chakrabarti:2006}. It may be noted that
this exponent value $\gamma=1$ is independent of the
dimension of the Cantor sets
 considered (here $\log 2/\log 3$) or for that
matter, independent of the fractals employed.
It also denotes the general validity of (\ref{bkc-eq:GR}) even for disordered
fractals, as observed numerically \cite{bkc-Pradhan:2003}.

Identifying the contact area or overlap $s$ between the self-similar
(fractal) crust and tectonic plate
 surfaces as the stored elastic energy $E$
released during the slip, the distribution (\ref{bkc-eq:GR}), of which a
derivation is partly indicated here,
reduces to the Gutenberg-Richter law (\ref{bkc-GR}) observed.

\section{Summary and Discussions}

Unlike the elastic constants (e.g., $Y$ in eqn. (\ref{bkc-sigma_f})) of a solid,
which can be estimated from the interatomic interactions and lattice
structures of a solid and while the effect of disorder on them can be 
accommodated using simple analytic formulas, the 
fracture strength ($\sigma_f$)
of a solid cannot be estimated easily from there and such estimates are
orders of magnitude higher than those observed. The reason is that cracks
nucleate around a defect or disorder in the solid and the variation of 
$\sigma_f$ with the defect size $l$ can be precisely estimated in a brittle
solid using the Griffith's formula (\ref{bkc-sigma_f}) in Sec. 2.1.

For disordered solids, the failure strength distribution $F(\sigma)$
depends on the sample volume and is given by the
 extreme statistics:
Gumbel\cite{bkc-Chakrabarti:1997} 
or Weibull 
\cite{bkc-Chakrabarti:1997} type
(see Fig. 2 for their generic behavior). The average
strength of a (finite volume) sample can be estimated from a modified
Griffith-like formula (\ref{bkc-T_f}). When percolation 
correlation length
exceeds the Lifshitz length, for large disorder, self-averaging statistics
takes over (see Sec. 2.2) and the average strength, given by (\ref{bkc-T_f}),
becomes precisely defined (even for infinite system size) and 
$\sigma_f$ becomes volume independent
(as in the equal load sharing fiber bundle model).

The inherent mean-field nature of the 
equal load sharing models 
(discussion in Sec. 2.3) enables us to construct recursion relations 
(Eq. (\ref{bkc-recU_t}) for example) which captures essentially
all the intriguing features of the failure dynamics. Though we have
identified 
$O\equiv U^{*}(\sigma )-U^{*}(\sigma_{f})\propto (\sigma_{f}-\sigma)^{\beta }$ 
as the order parameter (with exponent \( \beta =1/2 \)) for the continuous
transition in such models, unlike in the conventional phase transitions
it does not have a real-valued existence for \( \sigma >\sigma_{f} \).
The `type' of phase transition in such models has been a controversial
issue. Earlier it was suggested to be a first order
phase transition, because the the surviving fraction of fibers has
a discontinuity at the breakdown point of the bundles. However, as
the susceptibility shows divergence 
(\( \chi \propto (\sigma _{f}-\sigma )^{-\gamma };\gamma =1/2 \))
at the breakdown point, the transition has been later identified to
be of second order \cite{bkc-Pradhan:2001}. The dynamic critical
behavior of the these models and the 
universality of the exponent values
are straightforward. Here, divergence of relaxation time (\( \tau  \))
at the critical point 
(\( \tau \propto (\sigma _{f}-\sigma )^{-\alpha };\alpha =1/2 \))
indicates `critical slowing' of the dynamics which is characteristic
of conventional critical phenomena. At the critical point, one observes
power law decay of the surviving fraction in time 
(\( U_{t}(\sigma_{f})\propto t^{-\delta }; \)\( \delta =1 \)).
We demonstrated the universality of the failure behavior near 
\( \sigma =\sigma _{f} \),
for three different distributions: uniform, linearly increasing
and linearly decreasing distributions of fiber strength.
The critical strengths of the bundles differ in each case: 
\( \sigma _{f}=1/4,\sqrt{4/27} \)
and \( 4/27 \) respectively for these three distributions. However,
the critical behavior of the order parameter \( O \), susceptibility
\( \chi  \), relaxation time \( \tau  \) and of the time decay at
\( \sigma _{f} \), as given by the exponents \( \beta ,\gamma ,\alpha  \)
and \( \delta  \) remain unchanged: \( \alpha =1/2=\beta =\gamma  \)
and \( \delta =1 \) for all three distributions.

The model also shows realistic nonlinear deformation behavior
with a shifted (by \( \sigma _{L} \), away from the origin) uniform
distribution of fiber strengths (see Sec. 2.3.2). 
The stress-strain curve for the model
clearly shows three different regions: elastic or linear part (Hooke's
region) when none of the fibers break (\( U^{*}(\sigma )=1 \)), plastic
or nonlinear part due to the successive failure of the fibers 
(\( U^{*}(\sigma )<1 \))
and then finally the stress drops suddenly (due to the discontinuous
drop in the fraction of surviving fibers from \( U^{*}(\sigma _{f}) \)
to zero) at the failure point \( \sigma _{f}=1/[4(1-\sigma _{L})] \).
This nonlinearity in the response (stress-strain curve in Fig. 6) results
from the linear response of the surviving fibers who share the extra load
uniformly.
The local load sharing 
bundles (see Sec. 2.3.3) on the other hand show 
`zero' critical strength 
as the bundle size goes to infinity in one dimension 
(extreme statistics takes over). It is not clear at this stage if, in higher 
dimensions, LLS bundles are going to have non-zero critical strength.
In any case, the associated dynamics of failure of these higher dimensional 
bundles with variable range load transfer should be interesting.

We believe, the elegance and simplicity of the model, its common-sense
appeal, the exact solubility of its critical behavior in the mean
field (ELS) limit, its demonstrated 
universality, etc, would promote
the model eventually to a level competing with the Ising model of
magnetic critical behavior.

As emphasized already, we consider the physicist's identification
of the Gutenberg-Richter law (3) as 
an extremely significant 
one in geophysics. Like the previous attempts 
\cite{bkc-Burridge:1967,bkc-Carlson:1989,bkc-Bak:1997}, the model developed here
\cite{bkc-Chakrabarti:1999} captures this important feature in its
resulting statistics. 
Here, the established self-similarity of the fault planes are captured using
fractals, Cantor sets
 in particular. Hence we consider, 
in Sec. 3, this 
`Chakrabarti-Stinchcombe' model
 \cite{bkc-Bhattacharyya:2005}, where one
Cantor set moves uniformly over another similar set (with periodic boundary
conditions). The resulting overlap $s$ (meaning the set of real numbers
common in both the Cantor sets) changes with time: see, for example, 
Fig. \ref{bkc-timeseries} for a typical time variation overlap $s$ for $n=4$.
The number density of such overlaps seem to follow a Gutenberg-Richter 
type law (\ref{bkc-eq:GR}).
Judging from the comparisons
of the exponent values $\alpha$ in (\ref{bkc-GR}) and $\gamma$ in (\ref{bkc-eq:GR}),
the model succeeds  at least as well as
 the earlier ones. More
importantly, our  model incorporates 
both the geologically observed facts: fractal nature of the 
contact surfaces of the crust and of the tectonic plate,
and the stick-slip motion between them. However, the origin of the power
law in the quake statistics here is the self-similarity of the
fractal surfaces, and not any self-organisation directly in their 
dynamics. In fact, the extreme non-linearity in the nature of the
crack propagation is responsible for the fractal nature
of the rough crack surfaces of the crust and the 
tectonic plate. This
in turn leads  here to the Gutenberg-Richter like power law
  in
the earthquake statistics.

\medskip

\noindent
\textbf{Acknowledgments:} The author thanks M. Acharyya, 
K. K. Bardhan, L. G. Benguigui, 
P. Bhattacharyya, A. Chatterjee, D. Chowdhury, M. K. Dey, A. Hansen, 
S. S. Manna, S. Pradhan, P. Ray, D. Stauffer and R. B. Stinchcombe for 
collaborations at different stages.

\end{document}